\documentstyle[aps,prl,multicol,psfig]{revtex}
\begin{document}
\draft
\title{Unrestricted slave-boson mean-field approximation for the
       two-dimensional Hubbard model}
\author{G. Seibold and E. Sigmund}
\address{Institut f. Physik, Technische Universit\"at Cottbus,\\ 
        PBox 101344, 03013 Cottbus, Germany}
\author{V.~Hizhnyakov}
\address{Institute of Physics, Estonian Academy of Sciences, Riia 142,
        Tartu, Estonia}
\maketitle

\begin{abstract}
The Kotliar-Ruckenstein slave-boson scheme is
used to allow for an unrestricted variation of the bosonic and
fermionic fields on the saddle-point level. 
Various inhomogeneous
solutions, such as spin polarons and domain walls, 
are discussed within the two-dimensional Hubbard model and
compared with results of unrestricted Hartree-Fock (HF) calculations.
We find that the present approach drastically reduces the
polarization of these states and leads to increased delocalized 
wave functions as compared to the HF model.  
The interaction between two spin-polarons turns out to be attractive
over a wide range of the on-site repulsion U.
In addition we obtain the crossover from vertical to diagonal domain
walls at a higher value of U than predicted by HF. 
\end{abstract}

\begin{multicols}{2}  
\section{Introduction}
The unrestricted Hartree-Fock (HF) approach has turned out to be
a powerful tool in the calculation of inhomogeneous states
in the Hubbard model \cite{SU,POIL,SCHULTZ2,ZAANEN,INUI,BISHOP}. 
Among them magnetic polarons \cite{SU,SCHRIEFF,PHAS1,PROC,GOETZ}, domain walls 
\cite{POIL,SCHULTZ2,ZAANEN,INUI,SCHULTZ,OLES} and vortex solutions 
\cite{BISHOP}
have been extensively studied within the context of high T$_{c}$
superconductors (HTSC). 

Schrieffer, Wen and Zhang \cite{SCHRIEFF} have proposed the 
so-called spin-bag mechanism according to which a hole
couples to the spin density of the antiferromagnetically (AF)
ordered background of the CuO$_{2}$ planes thus creating
a local reduction of the AF order parameter. Within this model
two holes are attracted by sharing a common bag. In the simplest model
the resulting superconducting order parameter is of the order of
the spin-density wave gap which is large enough to
lead to high-temperature superconductivity.
The concept of spin polaron formation is also successful
in explaining the phase diagram in the low and intermediate
doping regime of the HTSC's. According to the concept of
microscopic electronic phase separation \cite{PHAS1,PROC},
the doping induced spin polarons (or spin clusters) form
a conducting subsystem in an background dominated by strong
antiferromagnetic correlations. Due to the competition between 
short-range attraction and the long-range
Coulomb repulsion, these polarons do not form a homogeneous sea in an
antiferromagnetic background.  Instead, the holes are
confined to a percolative network which dimension D is lower than
two ($1 \le D \le 2$). The resulting inhomogeneity in the electronic
subsystem has also important implications on the pairing mechansim
\cite{HIZ}.

Besides spin-polarons, domain wall solutions presently also attract a lot of
interest in the field of HTSC.
Stripe correlations have been observed in La$_{2-x}$Sr$_{x}$CuO$_{4}$ 
\cite{CHEONG,MASON,THURST}, in nickel oxide analogues of the
copper oxides \cite{TRAN} and in
La$_{1.6-x}$Nd$_{0.4}$Sr$_{x}$CuO$_{4}$ \cite{TRAN1,TRAN2}.
In the two latter systems the density fluctuations are pinned by
the underlying lattice structure giving rise to a static charge- and
spin-density wave.
The low-temperature orthorhombic phase in the nickel oxides stabilizes
the stripe structure along the diagonals whereas the low-temperature
tetragonal  structure in La$_{1.6-x}$Nd$_{0.4}$Sr$_{x}$CuO$_{4}$ favors the
pinning of a horizontal stripe phase.
Also in the CuO$_{2}$-planes of Bi2212 compounds
the existence of charge stripe order has been demonstrated 
with extended x-ray-absorption fine structure (EXAFS) experiments \cite{BIANC}.

The stripe instability was predicted theoretically in
\cite{ZAANEN} within a HF formalism applied
to the extended Hubbard model. Also most of the further investigations
on the striped phases \cite{POIL,INUI,SCHULTZ,OLES} have been carried out
within standard HF theory.
However, the solutions obtained with HF are poor variational
wave functions since they are much too high in energy.
Within HF, the only mechanism avoiding double occupancy in order
to reduce the Hubbard repulsion is to renormalize the
spin-dependent on-site energy $\epsilon_{i,\sigma}=U\langle n_{i,-\sigma}
\rangle$. Therefore the commensurate
antiferromagnetic phase displays as an alternating shift
of the spin-up and spin-down one-particle levels respectively, 
which overestimates by far the polarization of the AF order.

It is well known that a part of the correlations between electrons of
opposite spins can be accounted for by using the Gutzwiller projector 
\cite{GUTZ1,GUTZ2}. For the AF ordered system this approach leads to 
wave functions that are very close in energy to the
solution obtained by Quantum Monte Carlo simulations \cite{SHIBA,HIRSCH}.
In addition, the Gutzwiller wave function leads to a significantly
lower spin polarization in the intermediate U regime as compared
to the HF result \cite{SHIBA}.
The Gutzwiller approach has also been used to improve
the solutions of inhomogeneous states such as
spin polarons \cite{GOETZ,COPP} and domain walls \cite{COPP,GIAM}.
However, due to computional limitations these calculations where
done with an Ansatz for the charge- and spin-density profile, 
i.e the number of 
variational parameters for charge and spin at the different lattice sites
has been strongly reduced.

In this paper we overcome the limitations discussed above and present results
for unrestriced spin-polaronic and kink type Gutzwiller wave functions.   
Our approach is based on the representation of the Hubbard model
in terms of fermions and slave bosons due to Kotliar and
Ruckenstein (KR) \cite{KOTLIAR}. The KR formulation is a functional
integral method that reproduces the Gutzwiller solution at
the saddle-point level. The advantage of this method is that it
provides a systematic way to improve the solution by expanding
the fields around the saddle point. 
The KR approach is a good starting point for our purposes, since at the
saddle point level we can immediately identify the variational
parameters with the bosonic and fermionic fields.   
Since we don't make any assumption on the spatial symmetry of these
fields we therefore obtain various inhomogeneous textures which
stability depends on doping and the value of the Hubbard
repulsion U. In the present paper we will concentrate on the
stability and shape of spin-polaronic and domain wall solutions,
respectively. It turns out that for commensurate filling
(i.e. one hole along the wall) the parameter range for
the occurence of stripes is significantly enlarged in comparison
to the HF calculation. Within the investigated range of U
($4t \le U \le 10t$) we don't observe a crossover to a polaronic
Wigner crystal for commensurate doping, which means that the interaction 
between the holes keeps attractive up to very large values of U.     

The rest of the paper is organized as follows: In Sec. II we give a
detailed description of the formalism, in Sec. III we present
the results for spin polaronic and domain wall solutions respectively,
and in Sec. IV we summarize our conclusions.

\section{Model and Formalism}
We consider the two-dimensional Hubbard model on a square lattice, with 
hopping restricted to nearest neighbors (indicated by the bracket $<i,j>$)
\begin{equation}\label{HM}
H=-t\sum_{<ij>,\sigma}c_{i,\sigma}^{\dagger}c_{j,\sigma} + U\sum_{i}
n_{i,\uparrow}n_{i,\downarrow}
\end{equation}
where $c_{i,\sigma}^{(\dagger)}$ destroys (creates) an electron 
with spin $\sigma$ at site
i, and $n_{i,\sigma}=c_{i,\sigma}^{\dagger}c_{i,\sigma}$. U is the
on-site Hubbard repulsion and t the transfer parameter. For the calculations
in Sec. III we take t=1.
Following KR we enlarge the original Hilbert space by introducing
four subsidiary boson fields $e_{i}^{(\dagger)}$, 
$s_{i,\uparrow}^{(\dagger)}$, $s_{i,\downarrow}^{(\dagger)}$,
and $d_{i}^{(\dagger)}$ for each site i. 
These operators stand for the annihilation (creation) of
empty, singly occupied states with spin up or down, and doubly occupied 
sites, respectively. Since there are only four possible states per site, 
these boson projection operators must satisfy the completeness condition
\begin{equation}\label{CONST1}
e_{i}^{\dagger}e_{i}+\sum_{\sigma}s_{i,\sigma}^{\dagger}s_{i,\sigma}
+d_{i}^{\dagger}d_{i}=1
\end{equation}
Furthermore
\begin{equation}\label{CONST2}
n_{i,\sigma}=s_{i,\sigma}^{\dagger}s_{i,\sigma}+d_{i}^{\dagger}d_{i}
\end{equation}

\end{multicols}

Then, in the physical subspace defined by Eqs. (\ref{CONST1},\ref{CONST2})
the Hamiltonian (\ref{HM}) takes the form
\begin{eqnarray}
\tilde{H}&=& -t\sum_{<ij>,\sigma}z_{i,\sigma}^{\dagger}c_{i,\sigma}^{\dagger}
c_{j,\sigma}z_{j,\sigma} + U\sum_{i}d_{i}^{\dagger}d_{i} \label{SB}\\
z_{i,\sigma}&=&\frac{1}{\sqrt{e_{i}^{\dagger}e_{i}+s_{i,-\sigma}^{\dagger}
s_{i,-\sigma}}}(e_{i}^{\dagger}s_{i,\sigma}+s_{i,-\sigma}^{\dagger}d_{i})
\frac{1}{\sqrt{d_{i}^{\dagger}d_{i}+
s_{i,\sigma}^{\dagger}s_{i,\sigma}}}
\end{eqnarray}
and has the same matrix elements as those calculated for (\ref{HM}) in the
original Hilbert space.

\begin{multicols}{2}

In the saddle-point approximation, all bosonic operators are treated
as numbers. The resulting effective one-particle Hamiltonian 
describes the dynamics
of particles with modulated hopping amplitude and can be diagonalized
by the transformation
\begin{equation}
c_{i,\sigma}=\sum_{k}\Phi_{i,\sigma}(k)a_{k}
\end{equation}
where the orthogonality of the transformation requires 
\begin{equation}\label{CONST3}
\sum_{i,\sigma}\Phi^{\ast}_{i,\sigma}(k)\Phi_{i,\sigma}(q)=\delta_{kq}.
\end{equation}

Given a system with $N_{el}$ particles we finally obtain for the
total energy
\begin{equation}\label{E1}
E_{tot}=-t\sum_{<ij>,\sigma}z_{i,\sigma}^{\ast}z_{j,\sigma}\sum_{k=1}^{N_{el}}
\Phi^{\ast}_{i,\sigma}(k)\Phi_{j,\sigma}(k)+U\sum_{i}d_{i}^{2}
\end{equation}
which has to be evaluated within the constraints (\ref{CONST1},\ref{CONST2},
\ref{CONST3}).
This is achieved by adding these constraints quadratically to Eq. (\ref{E1})
\begin{mathletters}\label{C}
\begin{equation}
E_{C1}=\lambda_{1}\sum_{i}(e_{i}^{2}+\sum_{\sigma}s_{i,\sigma}^{2}
+d_{i}^{2}-1)^{2}\label{C1}
\end{equation}
\begin{equation}
E_{C2}=\lambda_{2}\sum_{i,\sigma}(\sum_{k}\Phi^{\ast}_{i,\sigma}(k)
\Phi_{i,\sigma}(k)-s_{i,\sigma}^{2}-d_{i}^{2})^{2}\label{C2}
\end{equation}
\begin{equation}
E_{C3}=\lambda_{3}\sum_{k,q}(\sum_{i,\sigma}\Phi^{\ast}_{i,\sigma}(k)
         \Phi_{i,\sigma}(q)-\delta_{kq})^2\label{C3}
\end{equation}
\begin{equation}
E_{C4}=\lambda_{4}(\sum_{k,i}\Phi^{\ast}_{i,\uparrow}(k)
         \Phi_{i,\uparrow}(k)-N_{\uparrow})^{2}\label{C4}
\end{equation}
\begin{equation}
E_{C5}=\lambda_{5}(\sum_{k,i}\Phi^{\ast}_{i,\downarrow}(k)
         \Phi_{i,\downarrow}(k)-N_{\downarrow})^{2}\label{C5}
\end{equation}
\end{mathletters}
We have added the last two conditions that turn
out to be very convenient because they allow to define the total number
of spin up and down particles $N_{\uparrow}+N_{\downarrow}=N_{el}$.
The energy functional $E\lbrace \Phi_{i,\sigma}(k),e_{i},s_{i,\sigma},
d_{i}\rbrace
=E_{tot}+E_{C1}+E_{C2}+E_{C3}+E_{C4}+E_{C5}$  now has to be minimized
with respect to the fermionic and bosonic fields. 
Since the KR theory does not preserve the spin-rotation 
invariance of the original hamiltonian all variational parameters
can be taken as real numbers \cite{COMMENT}.  
For the minimization procedure we have used a standard conjugate
gradient algorithm. The gradients of the functional $E\lbrace 
\Phi_{i,\sigma}(k),e_{i},s_{i,\sigma},d_{i}\rbrace$ can be calculated
analytically and convergence is checked by evaluating the norm
of the gradient. 
The accuracy of the solution can be controlled by calculating
the value of $E_{C1}+E_{C2}+E_{C3}+E_{C4}+E_{C5}$ at the end of
the iteration procedure. We generally have set the values
of the Lagrange parameters $\lambda_{1}...\lambda_{5}$ to $10^{4}-10^{5}$
which leads to an estimated Error at $\approx 0.0002$.

In principle one could start the calculation with a random configuration
of the fermionic and bosonic fields. However, for a doped system 
there exist different self-consistent solutions which are close
in energy and determining the most favorable can be difficult. 
Therefore, we have generally started from the 
unrestricted HF solutions for spin polaronic or domain wall
phases. The order of magnitude of the time needed to get 
convergence is half an hour on a SGI Indy workstation.

\section{Results}
In a first step we consider a single spin-polaron and compare the
spin and charge profiles obtained within the HF and unrestricted
SB approximation, respectively. We then extend the calculations to 
domain wall type solutions and study the stability of these
textures as a function of U. 
Before this, we evaluate the ground state energy and
AF polarization for the half filled system and compare our calculations
with the results of Yokoyama and Shiba \cite{SHIBA}. 
Since for this case their AF Gutzwiller variational approach (AFGF) is
equivalent to the saddle-point approximation of our slave boson
method we find perfect agreement within the numerical error.
In contrast to the HF scheme, the AFGF leads to a signicficant
reduction of the magnetization in the intermediate U regime.
On the other hand, both methods converge rapidly to fully polarized
S$_{z}$ components in the
large U-limit. One should note that Quantum Monte Carlo simulations 
\cite{HIRSCH} still lead to much smaller values in the large U limit. 

\subsection{Spin Polarons}
The formation of spin polarons in the 2D Hubbard model results 
from the competition between kinetic energy gain and 
magnetic energy loss when doping the system away from
half filling. Let us first consider the case where a particle with spin down
has been removed from the half filled, antiferromagnetically (AF) ordered 
lattice. If the vacancy is immobile, the cost in the
magnetic energy is 4J$\sim t^{2}/U$ in the large U limit. However, this
vacancy can gain kinetic energy via virtual hopping processes to the
nearest neighbor sites, thus mixing some probability of
spin up occupation to the site where the particle has been
removed from. Therefore, the resulting spin-density profile therefore
has an inverted order parameter at the site where the charge is
located. According to
Nagaoka's theorem \cite{NAG} the size of this ferromagnetic core 
is expected to be very large for large values of U.

Fig.\ \ref{fig1} show the charge and spin density profile of such a polaron
obtained by unrestricted HF calculations and with the present method, 
respectively.
The calculation was done on a $8\times8$ lattice with on-site
repulsion U=6t. In both methods, the doped hole is mainly localized 
at site (4,5),
But, whereas the charge at this site is reduced to $\langle n \rangle
=0.53$ within the HF approach, the unrestricted SB method gives a
value of $\langle n \rangle =0.73$  only. Moreover, the AF
order parameter $\Delta_{i}^{S}=(-1)^{i_{x}+i_{y}}S_{i}^{z}$ 
at the polaron center is much less
affected in the SB mean field (MF) treatment 
($\Delta_{i}^{S}=-0.07$) than in the case of unrestricted HF 
($\Delta_{i}^{S}=-0.21$). This discrepancy can be understood as follows.
The HF theory only renormalizes the spin dependent on-site
energies. The removal of a spin down particle at site i leads to a relaxation
of the spin up on-site level at this site. 
As a consequence the alternating
on-site level shift, describing the AF order, is changed at site i where
5 neighbored spin up states have now acquired nearly the same
energy. 
Thus there is a strong hybridization between the spin-up states and 
one obtaines a large value for the reversed spin order parameter 
at the central site i. By contrast the kinetic energy in the spin-down channel
between site i and its nearest neighbors is very much reduced since
the corresponding spin-down on-site level is pushed to a high
energy. Note that the HF method
always leads to a very large spin-polarization, since this is the only
way within this approximation to minimize the on-site repulsion.

Let us now consider the removal of a spin down particle at site i
when calculated within the SB aproximation. Then the spin-up 
hopping channel allows for the hybridization of neighboring
spin up states with site i, which is comparable to the HF approach.   
However, since we now have an additional variational parameter per site 
(i.e. the boson field $\lbrace d_{i}\rbrace$) double occupancy can be
minimized at site i, whithout very much reducing the
kinetic energy in the spin-down channel between site i and its 
nearest neighbors. In fact the reason why the Gutziller
wave function leads to a lower energy than HF theory is that there
one can minimize double occupancy while keeping the
kinetic energy at a higher value as compared to the HF approach.
In other words, the localization behavior is much more
pronounced in the HF treatment than within the SB formalism.

The energy difference of the solutions in Fig.\ \ref{fig1} (U=6t) 
is of the same 
order of magnitude than for the homogeneous AF 
solutions of Ref. \cite{SHIBA}. The total energy per site calculated
within the HF approximation is $E^{HF}=-0.607t$ and for the
SB method we get $E^{SB}=-0.645t$.
Fig.\ \ref{fig2} shows the order parameter $\Delta_{i}^{S}$ as a function of U 
at the perturbed sites and far from the spin polaron, 
evaluated with HF and SB approximations respectively.
It turns out that the SB method drastically reduces the polarization 
at the center of the polaron as compared to HF. 
As can be further seen on Fig.\ \ref{fig2}, this reduction of polarization
is much stronger than in the residual AF ordered plane.
Within the SB scheme
the spin polaron acquires a ferromagnetic core for values $U>5t$
whereas this limit in HF is already achieved for $U \approx 3.3t$.
  
Finally we evaluate the static interaction between two spin-polarons
which is of special importance with regard to the spin-bag model 
\cite{SCHRIEFF} and phase separation scenarios \cite{PHAS1} of the
high-T$_{c}$ superconductors. To calculate the binding energy
we consider two holes with opposite spins placed at two neighbored
sites and compare the energy of this configuration with the energy
of two infinitely separated spin-polarons. It should be noted that
a positive binding energy means attraction.
The results are plotted in Fig.\ \ref{fig3}. Within the HF approach
we obtain an interaction between the two polarons which is attractive 
up to U $\approx 6.5t$, in agreement with the results of Ref. \cite{SU}.
The binding energy displays a maximum at $U\approx 4.5t$.
However, within the SB approximation the parameter space of attraction
is considerably enlarged. The maximum of the binding energy,
which is approximately twice the value of the HF calculation, now occurs
at $U\approx 8.5t$. Unfortunately, convergence of our  variational approach
becomes very slow in the very large U-regime.
In this limit one has to increase very much the parameters $\lambda_{i}$
in eqs. (\ref{C}) in order to keep the constraint induced error within the
desired limits. This fact causes the difficulty in exploring the whole
parameter range of attraction in Fig.\ \ref{fig3}. 

\subsection{Domain wall solutions}
For intermediate values of the on-site Hubbard repulsion U 
Hartree-Fock theory predicts the existence of domain
wall solutions, where the doped charged carriers are
localized within a stripe in horizontal or diagonal
direction. This stripe separates two
AF ordered regions with opposite sign in the AF order parameter.
Within the HF approximation it was shown \cite{INUI} that there
is a transition from horizontal to diagonal stripes when
the ratio between on-site repulsion U and transfer integral t
exceeds the critical value of $U/t\approx 3.6$. From a constrained
Gutzwiller variation of hyperbolic-type domain walls it was concluded in
\cite{GIAM}, that this limit probably is shifted to much lower values.

In the following we will compare energies for diagonal and horizontal
stripes with the energy of isolated spin-polarons using the 
unrestricted SB scheme. The calculations are made
for different lattice sizes (typically $17\times 4$, $13\times 6$, 
9$\times$ 8) 
by applying appropriate boundary conditions 
for each domain wall type (see Fig.\ \ref{fig4}). 
The choice of the supercell is, of course, a delicate issue 
since periodic boundary conditions in principle 
require an 'uneven $\times$ uneven' lattice for a diagonal wall
and an 'uneven $\times$ even' lattice for a vertical stripe.
To avoid the comparison between different lattice sizes for
different domain wall types we have
choosen the 'shifted boundaries' shown in Fig.\ \ref{fig4}b.
This means that along the y-direction the supercells are 
shifted by the extend of the diagonal domain wall in x-direction.
Consider for example a diagonal domain wall on a $13\times 6$ lattice.  
Then the site (N$_{x}$,1) is
connected with site (N$_{x}+6$,6) for N$_{x}<8$ and with site
(N$_{x}$+6-13,6) for N$_{x}\ge 8$.
Calculating the energy per site for the half-filled AF ordered
system, we find that the two kinds of boundary conditions
in Fig.\ \ref{fig4} differ in the result by $0.1\%$ for U$\approx$ 3t.
This difference rapidly vanishes with increasing U. 
 
In Fig.\ \ref{fig5} we show the charge- and spin density profile of a vertical
domain wall calculated with unrestricted HF and SB approximations
respectively.
The doping corresponds to one hole per site in the domain wall.
As for the spin polarons studied in the previous section the SB 
result displays a charge- and spin profile that is 
considerably enlarged compared to the HF solution. This is in agreement with
the calculations of Ref. \cite{GIAM}. 
The dip in the HF charge profile is nearly twice the
value of the SB approximation.

Fig.\ \ref{fig6} shows the energy per hole as a function of U
for various textures such as domain walls and spin polarons.
We also have investigated the half-filled wall with the on-wall
quadrupling of the period which has been intensively
discussed by Zaanen and Ole\'{s} in Ref. \cite{OLES}.
The energy has been calculated in a standard way \cite{OLES}
by comparing the energy of each texture with the energy
of the homogeneous AF ordered state with the same number
of holes (compared to half filling).
In case of the diagonal wall we additionally have choosen
the same shift of the boundaries for the reference AF lattice.

Within the HF approximation we obtain a crossover from vertical
to diagonal domain walls for U$\approx 3.8t$ and a crossover
to isolated spin-polarons at U$\approx 8t$. 
The energy for half-filled walls is always higher than for
isolated polarons. These results are
in complete agreement with earlier HF studies of
the two-dimensional Hubbard model \cite{INUI,OLES}.
However, the range of
stability for the vertical stripe solution is considerably
enlarged in the unrestricted SB approximation where
we obtain the crossover at U$\approx 5.7t$
This result is supported by Lancos diagonalization studies of the tJ-model
\cite{MOREO} and Monte Carlo methods of the one-band Hubbard model
\cite{MOREO2,IMADA,FURUK}. These works report a shift of the
static spin structure factor peak in the vertical direction when
the charge density is reduced away from half-filling in agreement 
with our findings. 
Also recent studies of the 2D tJ model within a
density matrix renormalization group approach \cite{WHITE} are in
agreement with a vertical striped phase.
In addition we don't observe a crossover
to a spin-polaronic Wigner crystal for the considered range of U (U $\le 10t$).
Instead the energy of half-filled walls turns out to be lower than
the energy of spin polarons. 
In fact, since in the SB approach the range of attraction 
between the holes is considerably enhanced, these textures are 
energetically favored which have the holes placed nearby each other.
However, for very large values of U we again expect a decay of the
stripe into isolated polarons.

\section{Conclusions}
We have shown that the unrestricted SB saddle-point approximation
is a simple and powerful tool to improve inhomogeneous
solutions obtained by the HF method. It turns out that 
this approach leads to a strong reduction of the spin-polarization 
of these inhomogeneities in the intermediate U regime. 
The most relevant feature of the SB approximation, however, is
the considerably enlarged range of attraction between spin-polarons
in comparison with the HF method.
This result has also a strong impact on the spin-bag model of high
T$_{c}$ superconductivity, since HF theory has restricted the validity 
of this model to small values of the Hubbard repulsion U. 

Regarding the domain-wall phases, we find that the crossover from vertical
to horizontal stripes is shifted to higher values of the on-site
repulsion U than predicted by HF theory. Moreover we don't observe
a crossover to isolated polarons for U$\le10t$.
This result supports the description of 
domain wall structures in the
La$_{2}$NiO$_{4}$ compounds \cite{LITTLEWOOD} where it is generally argued
that the occurence of diagonal walls is a result of mean field theory.
Within the present approach the range of stability of diagonal walls
is considerably enhanced in comparison to the HF approximation, where
for U$\approx 8t$ the walls decay into isolated polarons.

\acknowledgments
Valuable discussions with W. Essl are gratefully acknowledged.
We also thank A. Bill for a critical reading of the manuscript.

\end{multicols}

\begin{figure}
\caption{Charge- ($\langle n_{i} \rangle$) and Spin- ($\Delta_{i}^{S}$) density
         profiles for a spin polaron on a $8\times 8$ lattice. a) HF
         approximation; b) SB approximation. The Hubbard on-site repulsion
         is $U=6t$.}
\label{fig1}
\end{figure}

\begin{figure}
\caption{Spin-density order parameter $\Delta_{i}^{S}$ at the
         center of the spin polaron (two lower curves) and at
         maximum distance from the polaron (two upper curves).
         Solid line: SB approximation; Dashed line: HF approximation.}
\label{fig2}
\end{figure}

\begin{figure}
\caption{Binding energy of a pair of polarons with opposite spins
         placed at neighbored sites on a $8 \times 8$ lattice.
         Dashed line: HF approximation; Solid line: SB approximation}
\label{fig3}
\end{figure}

\begin{figure}
\caption{Sketch showing the boundary conditions in y-direction 
         which have been applied to describe
         vertical (a) and diagonal (b) domain walls respectively.
         In x-direction we have taken periodic conditions.
         The line crossings correspond to the lattice sites and
         the dashed lattices mark the choosen periodicity in
         the y-direction.}
\label{fig4}
\end{figure}

\begin{figure}
\caption{Charge- (a) and Spin- (b) density
         profiles in x-direction for a vertical domain wall 
         on a $17\times 4$ lattice.
         The number of holes is 4 since the best energy is obtained when
         there is one hole per site in the wall. Solid lines: SB approximation;
         Dashed lines: HF approximation; Short dashed: $\cosh$- (for charge)
         and $\tanh$- (for spin) functional fit to the SB solution.}
\label{fig5}
\end{figure} 

\begin{figure} 
\caption{Binding energy per hole for vertical stripes (full lines),
         and diagonal stripes (dashed lines) with one
         hole per site along the wall, vertical half-filled
         walls (dotted lines) and isolated spin polarons (dashed-dotted
         lines). The binding energy is defined as the difference in
         energy between a given texture and the homogeneous AF ordered lattice
         with the same doping. The domain wall solutions have been calculated
         on a $9 \times 8$ lattice, polarons on a $8 \times 8$ lattice.
         a)HF approach;
         b)SB approximation.}
\label{fig6}
\end{figure}

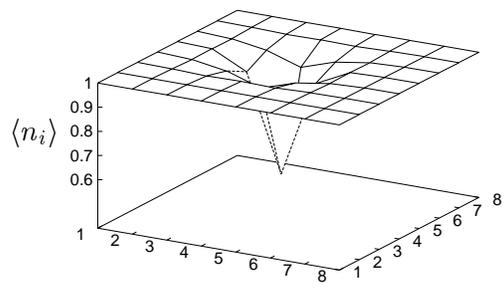
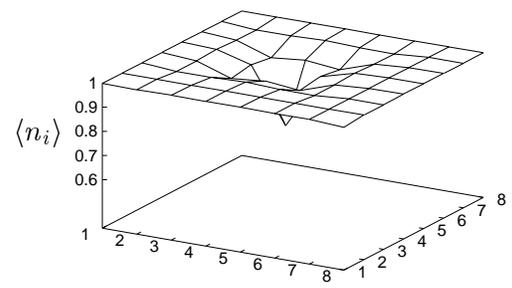
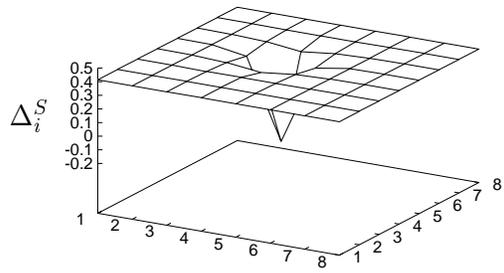
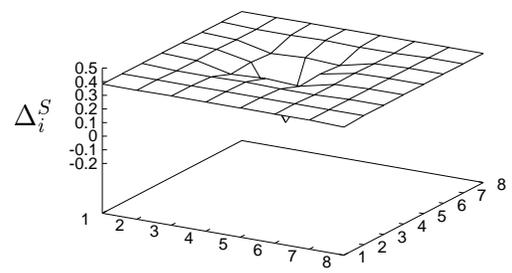
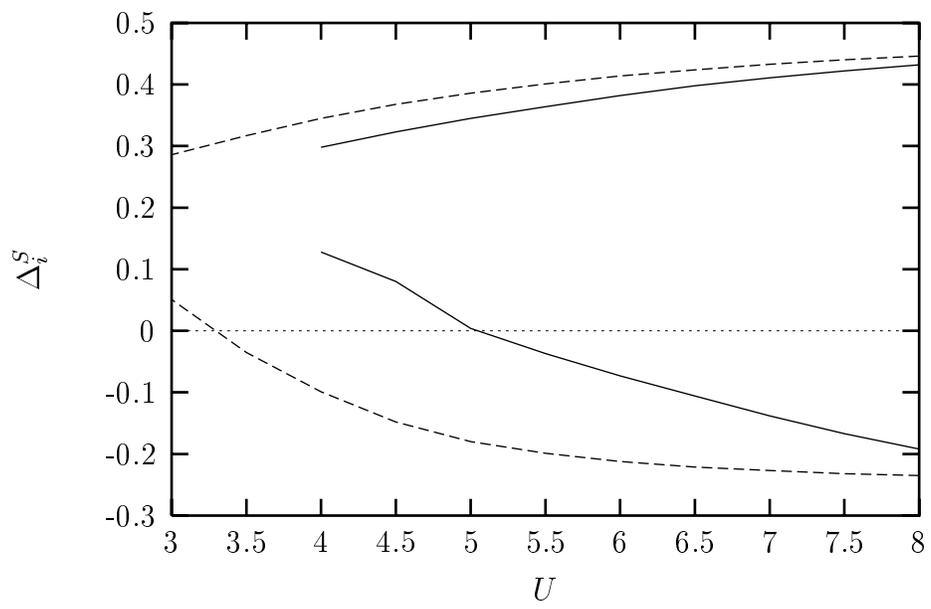
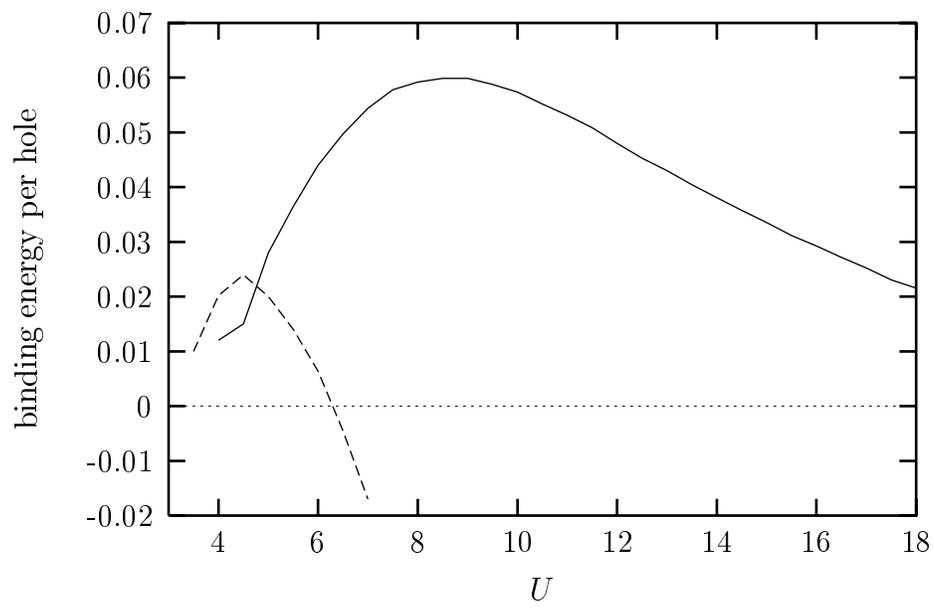
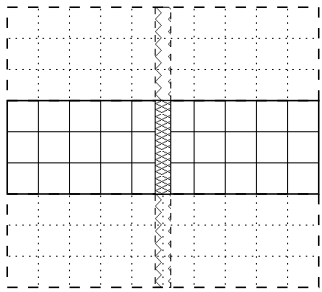
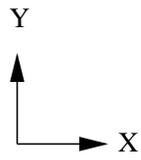
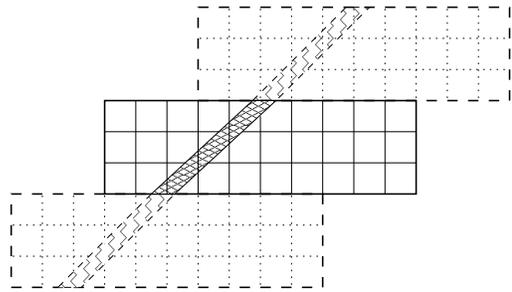
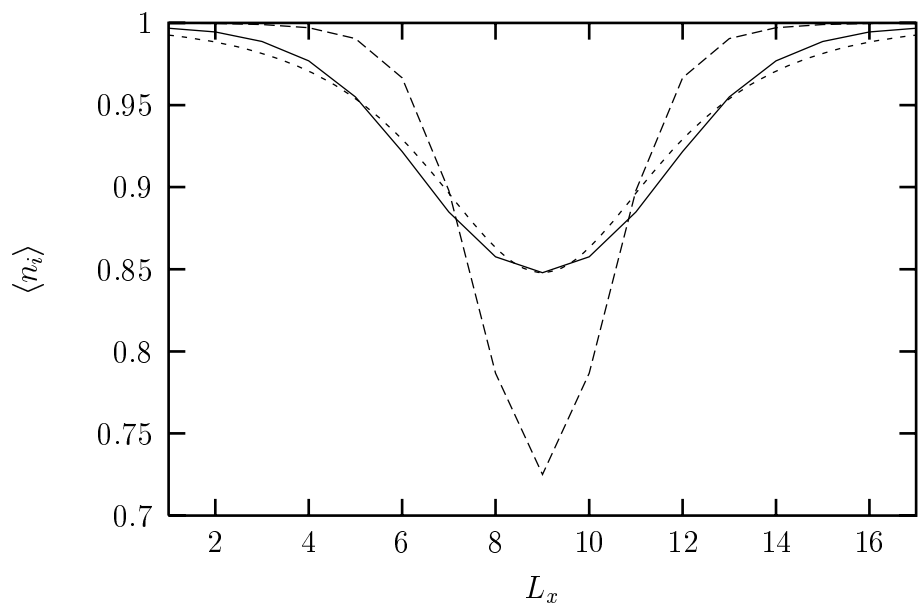
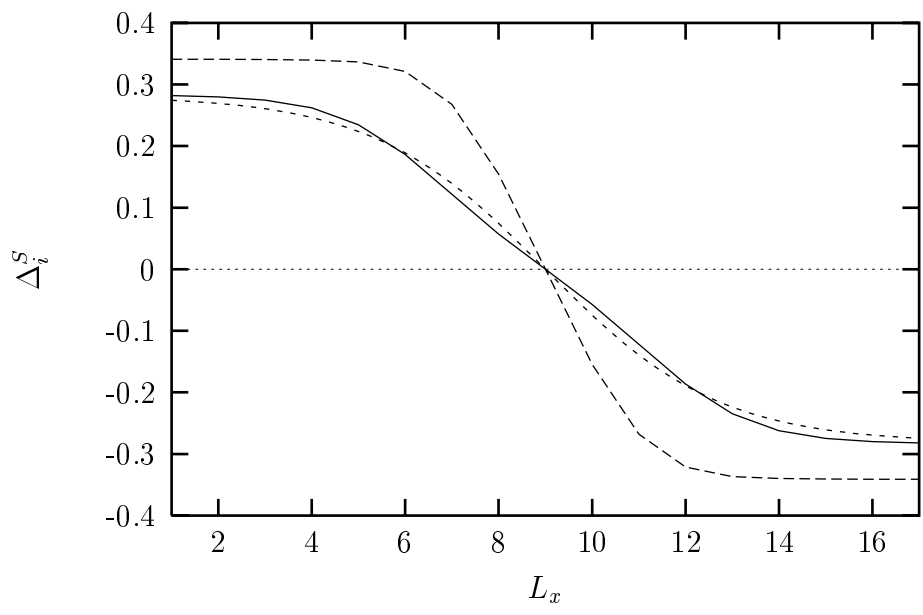
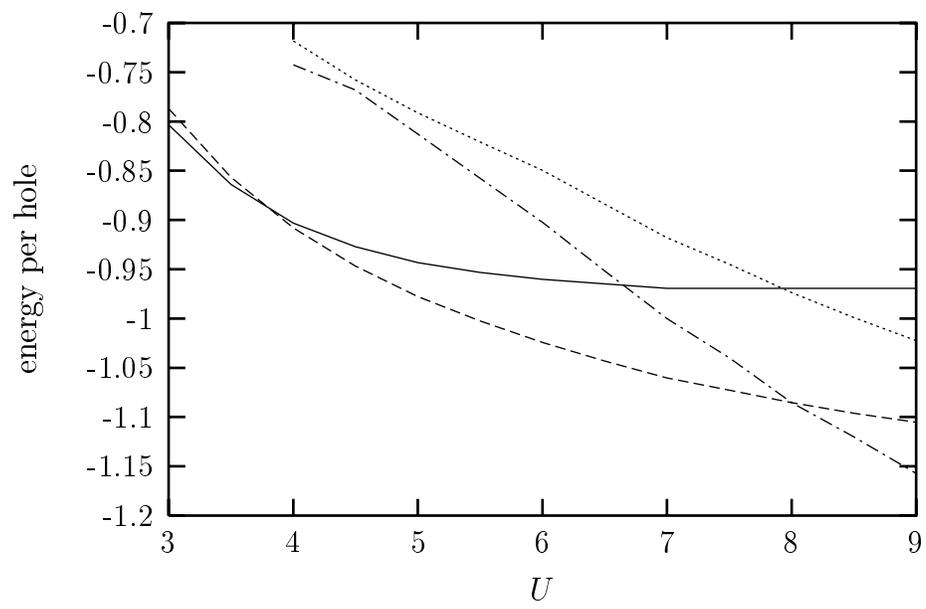
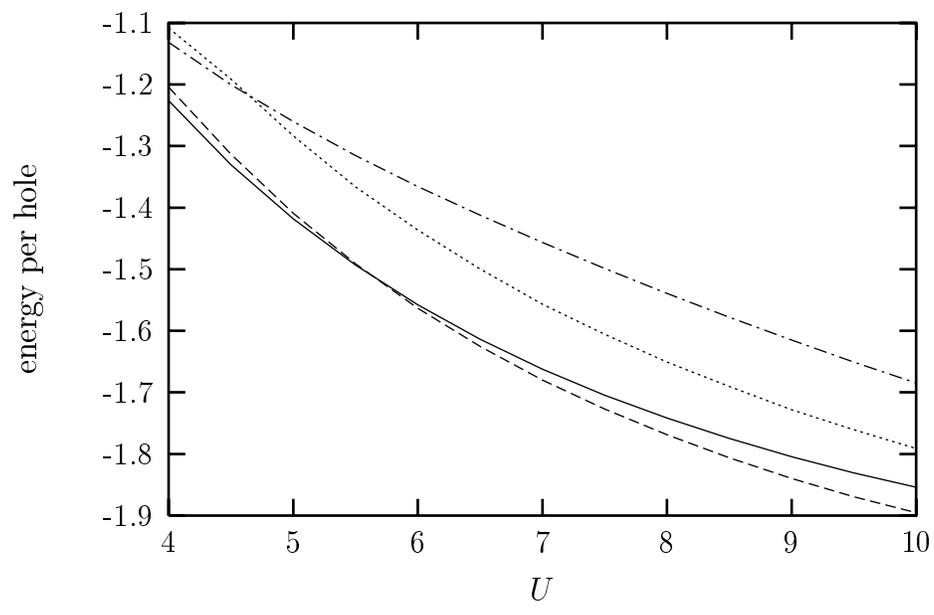
\end{document}